\documentclass[amsmath, amsfonts, superscriptaddress, showpacs, prl]{revtex4}
\usepackage{graphicx}
\usepackage{epsfig}
\usepackage{bm}
\usepackage{amsmath}
\usepackage{amssymb}
\usepackage{color}

\newcommand{\be}{\begin{equation}}
\newcommand{\ee}{\end{equation}}

\newcommand{\eps}{\epsilon}

\newcommand{\la}{\langle}
\newcommand{\ra}{\rangle}

\begin{document}

\title{Universal features of counting statistics of thermal
and quantum phase slips in superconducting nanocircuits}

\author{A.~Murphy}
\affiliation{Department of Physics, University of Illinois at
Urbana-Champaign, Urbana, Illinois 61801, USA}

\author{P.~Weinberg}
\affiliation{Department of Physics and Astronomy, Michigan State
University, East Lansing, Michigan 48824, USA}

\author{T.~Aref}
\altaffiliation[Present address:] {Department of Microtechnology and
Nanoscience (MC2), Chalmers University of Technology SE-412 96
G\"{o}teborg, Sweden, EU} \affiliation{Department of Physics,
University of Illinois at Urbana-Champaign, Urbana, Illinois 61801,
USA}

\author{U.C.~Coskun}
\affiliation{Department of Physics, University of Illinois at
Urbana-Champaign, Urbana, Illinois 61801, USA}

\author{V.~Vakaryuk}
\affiliation{Institute for Quantum Matter and Department of Physics
\& Astronomy, The Johns Hopkins University, Baltimore, Maryland
21218, USA}

\author{A.~Levchenko}
\affiliation{Department of Physics and Astronomy, Michigan State
University, East Lansing, Michigan 48824, USA}

\author{A.~Bezryadin}
\affiliation{Department of Physics, University of Illinois at
Urbana-Champaign, Urbana, Illinois 61801, USA}

\begin{abstract}
We perform measurements of phase-slip-induced switching current
events on different types of superconducting weak links and
systematically study statistical properties of the switching current
distributions. We employ two types of devices in which a weak link
is formed either by a superconducting nanowire or by a graphene
flake subject to proximity effect. We demonstrate that,
independently on the nature of the weak link, higher moments of the
distribution take universal values. In particular, the third moment
(skewness) of the distribution is close to $-1$ both in thermal and
quantum regimes. The fourth moment (kurtosis) also takes a universal
value close to $5$. The discovered universality of skewness and
kurtosis is confirmed by an analytical model. Our numerical analysis
shows that introduction of extraneous noise into the system leads to
significant deviations from the universal values. We suggest to use
the discovered universality of higher moments as a robust tool for
checking against undesirable effects on noise in various types of
measurements.
\end{abstract}

\date{May 21, 2013}


\maketitle

\twocolumngrid

\textit{Introduction}.--The field of quantum noise has recently seen
rapid development caused both by its growing significance in many
areas of condensed matter physics as well as by a constant
improvement in the capabilities of high precision
measurements~\cite{Review}. Perhaps the most intensively studied
question to date is related to the statistics of charge transport in
mesoscopic conductors. In such systems probability distribution of
current fluctuations, the so-called full counting statistics, was
rigorously derived for various normal and superconducting
circuits~\cite{Nazarov} and tested in the state-of-the-art
measurements of the third moment (skewness) of current
fluctuations~\cite{Prober}. Given that charge is a quantum
mechanical conjugate variable to the phase, it is of fundamental
interest to study corresponding statistics of phase fluctuations.
Superconducting nanowires and related proximity devices offer a
natural platform for this purpose, which we explore in the present
work.

The macroscopic quantum tunneling of the phase across a
current-biased Josephson junction~\cite{Devoret} or a
superconducting nanowire~\cite{Giordano,Bezryadin-PRL01,Sahu-NP09}
is arguably the most profound and well-known manifestation of
quantum fluctuations at the macroscopic level. This phenomenon is
observed by registering phase slip events~\cite{Little}, which
proliferate at currents close to the critical and drive transitions
between supercurrent-carrying and dissipative branches of
current-voltage
characteristics~\cite{Bezryadin-PRL01,Sahu-NP09,Aref-RPB12,Bezryadin-Book}.
Macroscopic quantum tunneling is usually described in the framework
that treats quantum or thermally activated transitions of the phase
between neighboring minima of a tilted washboard potential in the
presence of a dissipative environment~\cite{CL-model,
Grabert,Fisher}. Complimentary approaches employ an effective action
for BCS superconductors~\cite{GZ,Zaikin-Review,AL-PRB07}.

Unlike experiments associated with the charge transfer where
measurements of each moment of the full counting statistics is beyond current
experimental capabilities, experiments on switching current allow
one to reconstruct full distribution of superconducting phase
fluctuations since a single phase slip is sufficient to drive the
system into resistive state~\cite{Sahu-NP09,Gleb-PRL11} by creating
a hot spot~\cite{Shah-PRL07,Pekker-PRB09}. Thus there exists a
one-to-one correspondence between phase fluctuations - \textit{phase
slips} - and switching events.

In this Letter we report a systematic study of higher moments of the
switching current distribution as a function of temperature and
other parameters of our devices. The higher moments under
investigation include skewness $S$ that quantifies an asymmetry of
the distribution, and kurtosis  $K$ that is a measure of its
peakedness (for definitions see below). We present evidence, both
experimental and theoretical, that these higher moments are in fact
universal constants: $S\approx -1$ and $K\approx 5$. Surprisingly,
the observed crossover from a classical escape mechanism (i.e., the
thermal activation) to a quantum one (i.e., quantum tunneling from a
metastable energy minimum) does not lead to any noticeable changes
in these moments. We evince this universality using two types of
samples, namely graphene junctions under the proximity effect as
well as ultra-thin superconducting nanowires. Apparent universality
of $S$ and $K$ has to be contrasted with the behavior of the
standard deviation of the switching current (the second moment $\sigma$) that exhibits
nontrivial temperature dependence: the power-law~\cite{Kurkijarvi},
$\sigma\propto T^{2/3}$, in the thermal regime and $\sigma\propto
const$ in the quantum
regime~\cite{Sahu-NP09,Gleb-PRL11,Aref-RPB12,Bezryadin-Book}.

\textit{Devices}.--Nanowire samples were
prepared~\cite{Bezryadin-Nature-2000,Bezryadin-Book} by depositing
carbon nanotubes across a 100~nm wide trench on a silicon chip,
coated by a film of SiO$_2$ and a film of SiN. A film of 10-20nm of
Mo$_{76}$Ge$_{24}$ was sputtered onto the chip, covering the top SiN
surface and the nanotubes crossing the trench. Thus the suspended
segments of nanotubes were converted into nanowires. Uniform wires
were selected using SEM, and the MoGe film was patterned by
photolithography, to define contact pads (electrodes). In such
devices the selected nanowire serves as the only conducting link
connecting the superconducting thin-film electrodes, positioned on
the opposite sides of the trench. Importantly, there is no
additional contact resistance between the nanowire and the contact
pad since the wire transforms seamlessly into the pad while both are
made in the same sputtering run.

Graphene flakes were deposited onto SiO$_2$ surface by mechanical
exfoliation~\cite{Novoselov}. Electron-beam lithography was utilized to
pattern the electrodes into a comb shape. After the resist was
exposed and developed, we deposit, using thermal evaporation,
a 4~nm Pd film (so-called sticking layer) and a 100~nm Pb film on
the top. Lift-off was performed by placing the sample in an acetone
bath for five minutes, sonicating it for ten seconds every other
minute. The 100nm Pb layer induces superconductivity in the graphene
through the proximity effect. The samples were measured in a He-3
cryostat. Electromagnetic noise was filtered from the system using
$\pi-$filters at room temperature and a copper powder and
silver-paste radio-frequency noise filters at low temperatures.

A sinusoidal bias current, having an amplitude greater than the
critical current of the device, was applied across each sample. As
the current increased from zero to its maximum, the voltage across
the sample demonstrated a sudden jump from zero to some large,
non-zero value, indicating the system switched from a
superconducting state to a normal, resistive state. The value of
bias current at which the jump took place was recorded as the
switching current. Then the bias current returned to zero, and the
system once again became superconducting. This process was repeated
$N=10^4$ times (or 5000 in some cases) for each set of parameters. Each
measurement gave slightly different value of the switching current,
due to inherent stochasticity of the phase slips, thus producing
switching current distributions. The skewness and kurtosis of each
distribution was calculated from the recorded data by using standard
expressions: $S= N^{-1} \sum\limits_{i=1}^{N} (I_{sw,i}-\langle
I_{sw}\rangle)^3/\sigma^3$ and $K= N^{-1}\sum\limits_{i=1}^{N}
(I_{sw,i}-\langle I_{sw}\rangle)^4/\sigma^4$, where each $I_{sw,i}$
represents an applied  bias current at which a switching event took
place, $\langle I_{sw}\rangle$ is the mean switching current, and
$\sigma$ is the standard deviation of the switching distribution.

\begin{figure}[!]
 \includegraphics[width=9cm]{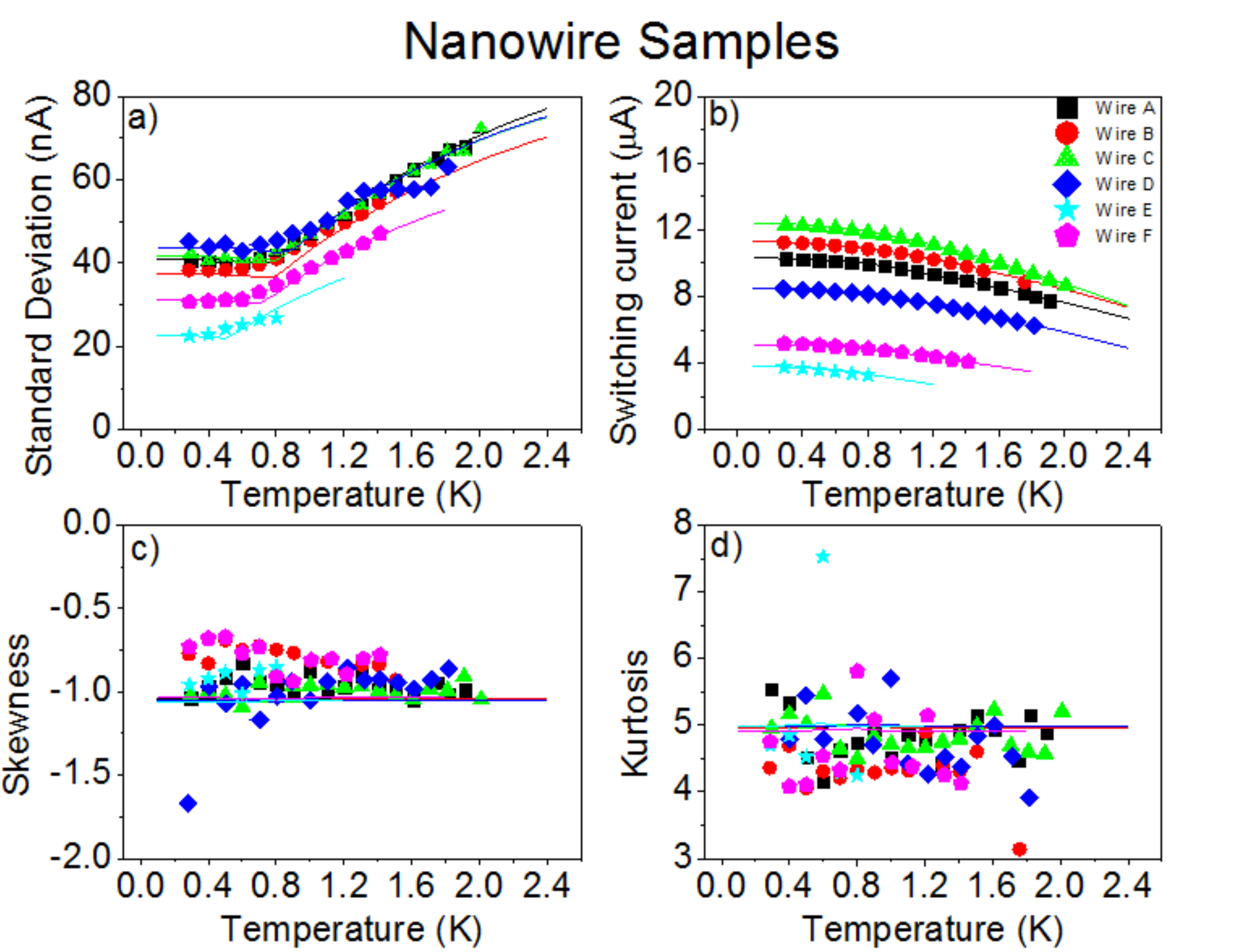}
  \caption{(a) standard deviation, (b) mean switching current,
  (c) skewness, and (d) kurtosis of the switching current distributions
  in nanowire samples A, B, C, D, E and F vs temperature $T$.
  The experimental values are represented by symbols.
  Simulation curves are shown by solid lines.
  One point in figure (c) and two points in figure (d) lie outside the ranges shown.
  Fitting parameters used in the simulation are summarized in Table I
  of the Supplementary Material~\cite{SM}.}\label{Fig Nanowire_All}
  \vspace{-10pt}
\end{figure}

\begin{figure}[!]
 \includegraphics[width=9cm]{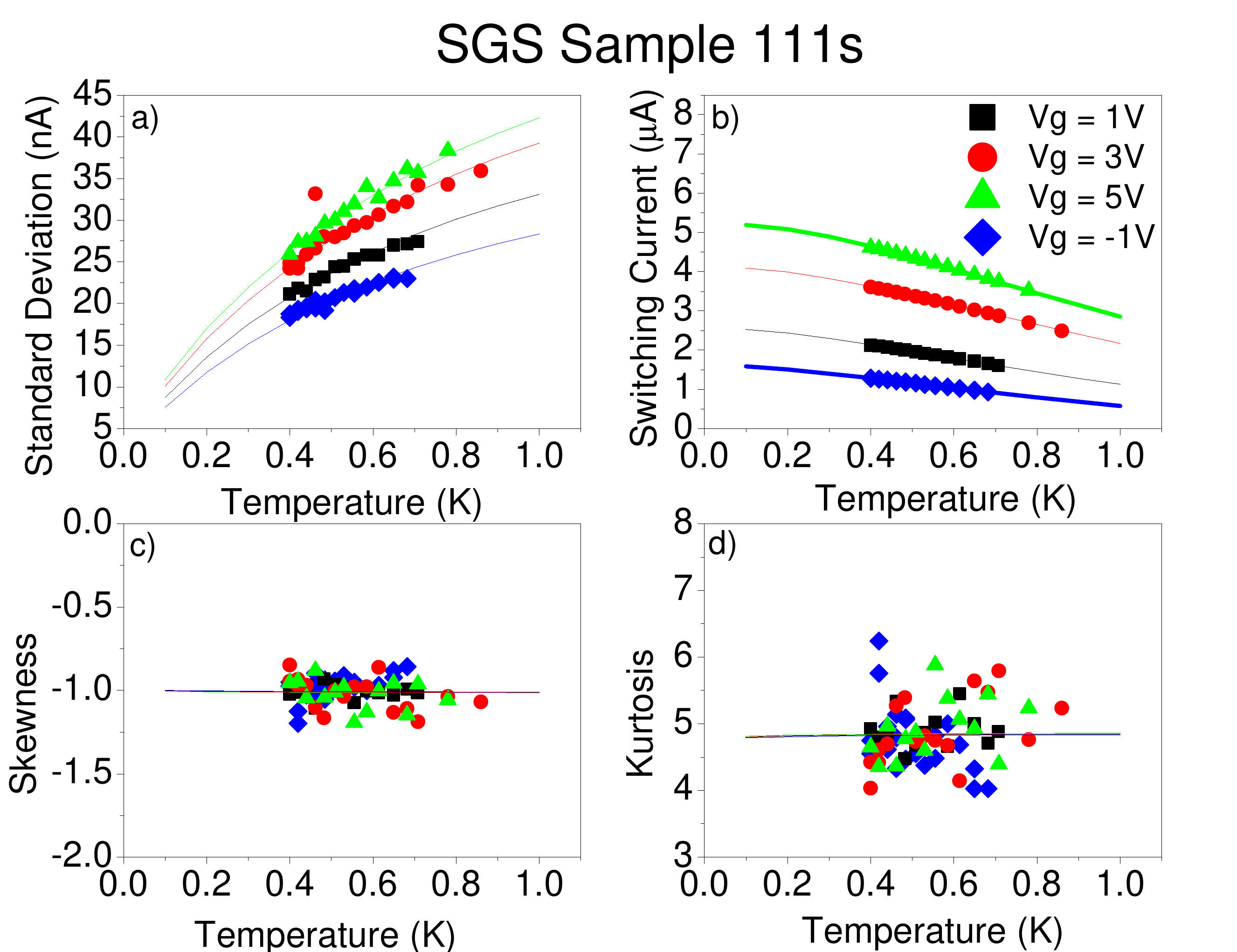}
 \caption{(a) standard deviation, (b) mean switching current,
 (c) skewness, and (d) kurtosis of the switching current distributions
 in SGS sample 111s vs temperature at a gate voltages of 1V, 3V, 5V and -1V.
 We use the same convention as in the previous figure.
 One point in figure (c) lies outside the range shown. Fitting parameters used in the simulation
 are summarized in Table II of the Supplementary Material~\cite{SM}.
 Data for SGS sample 105s are shown in~\cite{SM}.}  \label{Graphene_All}
 \vspace{-10pt}
\end{figure}

\textit{Experimental results}.--We first discuss the effect of
temperature on the skewness and kurtosis of the switching current
distributions. We find that in both types of samples -- superconducting nanowires
(Figs.~\ref{Fig Nanowire_All}c-\ref{Fig Nanowire_All}d), and
graphene proximity junctions
(Figs.~\ref{Graphene_All}c-\ref{Graphene_All}d) -- the skewness and
kurtosis are constant with temperature. Surprisingly, these moments
are identical within experimental uncertainty for the two
qualitatively different systems. The value of the skewness in both
cases is near $-1$, and the value of kurtosis is near $5$. In
nanowire samples, these moments remain constant even as the system
experiences a crossover from the high temperature regime, at which
phase slips are predominantly caused by thermally activation, to low
temperatures, at which quantum tunneling of phase slips is
responsible for the premature switching. This crossover is evident
in Fig.~\ref{Fig Nanowire_All}a as the standard deviation changes
from the power-law at high temperatures to a constant value at low
temperatures. The classical-to-quantum crossover temperature is
typically in the range 0.6-0.8~K for the studied samples
(Fig.~\ref{Fig Nanowire_All}).

In SGS samples in addition to the temperature dependence we also
study the effect of gate voltage $V_g$ on the skewness and kurtosis.
Both moments remain constant within
the experimental uncertainty over a wide range of $T$ and $V_g$ (see
Figs.~\ref{Graphene_All}c-d and \cite{SM}). It should be
noted that, unlike nanowire samples, SGS junctions do not show crossover
to the quantum tunneling dominated regime within experimentally
tested temperatures. However, we do expect that such crossover might
occur at lower temperatures, as recently reported~\cite{Lee-PRL11}.

We also demonstrate numerically~\cite{SM}, that the presence of
extraneous noise leads to a substantial reduction of the universal
moments. This observation provides an independent tool for assessing the
relevance of noise to  the interpretation of experimental data.

\textit{Numerical simulations and fitting}.-- The fitting curves
presented in Figs.~\ref{Fig Nanowire_All} and \ref{Graphene_All}
were obtained using the Arrhenius-type activation formula for the rate
of phase slips (hereafter
$\hbar=k_B=1$)~\cite{Tinkham,Bezryadin-Book}
\begin{equation}\label{rate}
\Gamma (I,T)=\Omega(I,T) \left[e^{-U(I,T)/T}+e^{-
U(I,T)/T_q}\right],
\end{equation}
which accounts for both thermal and quantum escape processes. Here
$\Omega$ is the attempt frequency, $U$ is the energy barrier for a
phase slip, $T$ is the base temperature and $T_q$ is the quantum
temperature used to model the regime of macroscopic quantum
tunneling observed in nanowire samples at low
temperatures~\cite{Note-1}. For both systems activation energy has
power-law functional dependence on the applied bias current
\begin{equation}\label{U}
U(I,T)=\frac{\kappa I_c(T)}{e}(1-I/I_c(T))^{\eta}.
\end{equation}

For SGS devices we took $\kappa=\sqrt{8}/3$ and
$\eta=3/2$~\cite{Coskun-PRL2012,Note-2} and used a critical current in
the form
\begin{equation}\label{Ic-SGS}
I_c(T)=\frac{64\pi T}{eR_N}\displaystyle
    \sum\limits_{n=0}^{\infty}
    \frac{\Delta^2 (L/L_n)\exp(-L/L_n)}
    {[\omega_n+W_n + \sqrt{2(W^2_n +\omega_nW_n)}]^2}
\end{equation}
where $R_N$ is the normal state resistance of a junction, $\Delta$
is the superconducting gap in the leads, $\omega_n=(2n+1)\pi T$,
$W_n=\sqrt{\Delta^2+\omega^2_n}$, $L_n=\sqrt{D/2\omega_n}$. The sum
over $n$ was taken until convergence (roughly 10 terms). Expression
\eqref{Ic-SGS} follows from the theory of disordered
superconductor-normal metal-superconductor
junctions~\cite{ZZ,Dubos-PRB2001,AL-PRB06}. It has to be stressed that
ballistic theory of the proximity effect in SGS
junctions~\cite{Titov} fails to account for the temperature and gate
voltage dependencies of the critical current for our devices (see
Fig.~\ref{Graphene_All}b). This observation is also consistent with
the previous reports on the proximity effect in SGS
systems~\cite{Lee-PRL11,Coskun-PRL2012,SGS-1,SGS-2,SGS-3,SGS-4,SGS-5,SGS-6}.
From the normal state resistance of our samples, we deduce typical
mean free path $l\sim20$nm, which correspond to the diffusion
coefficient $D\sim 50$cm$^2$/s. Because the mean free path and the Thouless
energy $E_{Th}=D/L^2\sim80\mu$eV are much smaller than the junction
spacing of $L\sim300$nm and the energy gap $\Delta\sim1$meV,
respectively, our SGS junctions correspond to a long diffusive
junction limit.

For superconducting nanowires there are two known models for $U$ in
Eq.~\eqref{U}. If a wire forms a phase slip junction (PSJ) then
$\kappa=\sqrt{6}/2$ and $\eta=5/4$~\cite{GZ,LA,MH,Tinkham-PRB}. The
corresponding expressions for $\kappa$ and $\eta$ for the more
thoroughly studied case of a Josephson junction have the same
values as above for the SGS devices. It is worth noting that
qualitatively the two models are very similar. Following the
previous work~\cite{Bardeentest} we model the critical current of nanowire
devices by the phenomenological Bardeen's formula~\cite{Bardeen}
\begin{equation}\label{Ic-NW}
I_c(T)=I_c(0)(1-T^2/T^2_c)^{3/2}.
\end{equation}
Finally, for both SGS and nanowire systems the escape attempt frequency in
Eq.~\eqref{rate} was described by
\begin{equation}\label{attempt-f}
\Omega(I,T)=\Omega_0(T)(1-I/I_c(T))^{\nu}
\end{equation}
with $\nu=1/4$ for JJ model, and $\nu=5/8$ for PSJ model.

Eqns.~\eqref{rate}-\eqref{attempt-f} were combined to determine the
rate of phase slips. For a given set of parameters this rate was
used to predict the switching distribution as a function of bias
current and then calculate its mean, standard deviation, skewness
and kurtosis. Such procedure was repeated at different temperatures
to produce the temperature dependence of the moments. In the case of
SGS samples the above scheme was also repeated at different gate
voltages. Parameters ($\Omega_0$, $I_c(0)$, $T_c$, $T_q$) for nanowire
samples and ($\Omega_0$, $R_N$, $T_c$ and $D$) for SGS samples were
then adjusted within the expected range of values until the
predicted switching current and standard deviation vs temperature
curves matched the data. These, along with the resulting skewness
and kurtosis curves were used as fits to the data and are plotted as
solid lines in Figs.~\ref{Fig Nanowire_All} and
\ref{Graphene_All}~\cite{SM,Note-3}.

\textit{Analytical model}.--In this section we compute skewness and
kurtosis by using an approach developed for the problem of escape
from a metastable potential well subject to a steadily increasing
bias field~\cite{Kurkijarvi,Garg-PRB1995}. We consider a general
situation in which the phase slip rate of the system -- either
thermal or quantum --  can be written in terms of the reduced
current variable $\eps=1- I/I_c$ as
\begin{equation}
\Gamma(\eps)=A\eps^{a+b-1}\exp (-B\eps^b).
\end{equation}
This form is general enough to cover all range of parameters relevant
for our experiment on both types of devices. The powers $a$ and $b$
depend on whether the escape is quantum or thermally activated,
while parameters $A$ and $B$ depend on the degree and type of
damping (in particular, we estimate that our SGS junctions are
moderately underdamped with the quality factor $Q\simeq4$). The
distribution function for phase slips can be expressed in
terms of the rate as
\begin{equation}
    P(\eps) = \frac 1 {|\dot \eps | } \Gamma (\eps)
     \exp\left[ - \frac 1 {|\dot \eps|} \, \int_\eps^\infty \Gamma(\eps') \,
     d\eps'\right]
     \label{p}
\end{equation}
where $|\dot \eps |$ is  a constant ramp speed. We are interested in
central moments $m_n$ of variable $\eps$ i.e.~moments defined around
its mean value $\bar \eps$:
\begin{equation}
    m_n \equiv \la \, (\eps - \bar \eps)^n \, \ra = \int_0^\infty d\eps \, (\eps - \bar \eps)^n\,  P(\eps)
\end{equation}
where $\bar \eps =  \int_0^\infty d\eps \, \eps \,P(\eps)$.
Dispersion, skewness and kurtosis can be expressed in terms of
central moments. To this end, it is convenient to introduce a
dimensionless parameter
\begin{equation}
    Z=\ln\left[\frac {A/|\dot \eps|}{ b B^{1+a/b}}\right],
\end{equation}
which only weakly depends on the characteristics of the system in
question. For self-consistency of the description this parameter
should be large which can be achieved by tuning the ramp speed
$|\dot \eps|$.

It is straightforward to show that moments of distribution (\ref{p})
can be written as an asymptotic power series in $1/Z\ll1$ as
follows:
\begin{equation}
    \la \, \eps^n \, \ra =
    B^{-n/b} Z^{n/b}
    \left[ 1+\sum_{j=1} Z^{-j} f_j(n, \ln Z)  \right].
    \label{epsn}
\end{equation}
Definition of the expansion coefficients $f_j$ are relegated to the
Supplementary Material~\cite{SM} because of their cumbersome form.
Within the model $f_j$ depend on power exponents $a$ and $b$, and
very weakly (as a double logarithm), on temperature-dependent
parameters $A$ and $B$, and the ramp speed $|\dot \eps|$. This
implies that temperature scaling of both moments $\la \, \eps^n \,
\ra$ and central moments $\la (\eps - \bar \eps)^n\ra$ is fully
dominated by the temperature scaling of $B$ which is proportional to
the height of the phase slip barrier $
    \{\la \, \eps^n \, \ra, \la (\eps - \bar \eps)^n\ra\} \propto
    B^{-n/b}(T)$.
To determine proportionality coefficients one needs to use the
explicit form of $f_j$. Up to the order $1/Z$ first two moments are
given by
\begin{eqnarray}
     &&\hskip-.55cm\bar \eps = Z^{1/b} B^{-1/b}
    \left(1 + \frac {v/b} {Z}\right),
    \\
     &&\hskip-.55cm\sigma^2   \equiv  m_2 =  Z^{2/b -2} B^{-2/b}
    \nonumber
    \\
     &&\hskip-.55cm\left( \frac {\pi^2}{6
     b^2}+\frac 1 {Z b^3}\big[a \pi^2/3 + (1-b) (\pi^2 v /3 -
     \psi''(1))\big]\right).
    \label{m2}
\end{eqnarray}
We have defined $v = (a/b) \ln Z +\gamma$, where $\gamma\approx
0.577$ is the Euler-Masheroni constant, and $\psi''(1)\approx-2.404$
is the tetragamma function~\cite{Note-4}. Despite increasing
complexity of the calculation the leading term in the third and
fourth central moments are given by simple expressions:
\begin{gather}
    m_3 =  B^{-3/b} Z^{3/b-3} \left(-\frac{\psi''(1)}{b^3} + \delta_3
    \right),
    \label{m3}
    \\
    m_4=  B^{-4/b} Z^{4/b-4} \left( \frac {3\pi^4}{20b^4} +
    \delta_4\right).
    \label{m4}
\end{gather}
The correction terms are of the order $\{\delta_3,\delta_4\}\propto
Z^{-1}$. For example
$\delta_3=\frac{1}{60Zb^4}[90a\pi^2v(v-1)-11\pi^4(b-1)-180\psi''(1)(a-v(b-1))]$~\cite{SM}.
We are now in the position to compute skewness and kurtosis and thus
find:
\begin{eqnarray}
    &&S=-m_3/m_2^{3/2}=6^{3/2}\psi''(1)/\pi^3+O(Z^{-1}),\\
    &&K=m_4/m_2^{2}=27/5+O(Z^{-1}),
\end{eqnarray}
which are central results of this section. Remarkably, to the
leading order in $1/Z\ll1$, both skewness and kurtosis are given by
universal numbers $S\approx-1.139$ and $K\approx5.4$, which are
independent of the parameters of the system and are the same for
both thermal and quantum phase slips. The magnitude of the
correction terms $\delta_3,\delta_4$ is analyzed for different
values of the ramp speed and different models of a weak link in~\cite{SM}.

\textit{Conclusion}.--We have experimentally demonstrated the
universality of higher moments -- skewness and kurtosis -- of the
switching current distribution in superconducting nanocircuits. Our
results are supported both by analytical modeling and by numerical
simulations. We have also pointed out that the universality of
higher moments is affected by extraneous noise~\cite{SM} and suggested to use
this observation to detect the presence of unwanted noise in the
data.

The work was supported by DOE DE-FG02-07ER46453, ONR N000140910689,
and NSF DMR 10-05645. V.~V. was supported by the U.S. Department of
Energy, Office of Basic Energy Sciences, Division of Materials
Sciences and Engineering under Award DE-FG02-08ER46544 and by the
Theoretical Interdisciplinary Physics and Astrophysics Center at
JHU. A.~L.~acknowledges support from Michigan State University.


\clearpage
\onecolumngrid
\setcounter{figure}{0}
\setcounter{equation}{0}

\section{Supplementary Material}

\subsection{Expansion Coefficients $f_j(n,\ln Z)$}

Introducing new variables $r=n/b$ and $v=(a/b) \ln Z +\gamma$, we
find that expansion coefficients introduced in Eq.~(10) of the main
text  are given by the following expressions:
\begin{widetext}
\begin{eqnarray}
&&f_1(n,\ln Z)=vr, \\
&&f_2(n,\ln Z)=\frac{a}{b}(v+1)r+\frac{r(r-1)}{2}\left(v^2+\frac{\pi^2}{6}\right),\\
&&f_3(n,\ln
Z)=-\frac{ar}{2b}\left(v^2+\frac{1}{b}[2v(b-a)+2b-3a]+\frac{\pi^2}{6}\right)+\frac{ar(r-1)}{b}
\left(v^2+v+\frac{\pi^2}{6}\right)\nonumber\\&&+\frac{r(r-1)(r-2)}{6}\left(v^3+\frac{\pi^2v}{2}-\psi''(1)\right),\\
&&f_4(n,\ln
Z)=\frac{ar}{3b}\left(v^3+\frac{\pi^2v^2}{2}-\psi''(1)+\frac{3(2b-3a)}{2b}\left(v^2+\frac{\pi^2}{6}\right)
-\frac{3av(3b-a)}{b^2}+\frac{3(2b-a)}{b}(v^2-\gamma+\gamma^2)\right)\nonumber\\
&&+\frac{r(r-1)}{2}\left(\frac{a^2}{b^2}\left((v+1)^2+\frac{\pi^2}{6}\right)-
\frac{a}{2b}\left(v^3+\frac{\pi^2v}{2}-\psi''(1)+\frac{2(b-a)}{b}
\left(v^2+\frac{\pi^2}{6}\right)+\frac{v(2b-3a)}{b}\right)\right)\nonumber\\
&&+\frac{ar(r-1)(r-2)}{2b}\left(v^3+v^3\frac{\pi^2+2}{2}-\psi''(1)+\frac{\pi^2}{6}\right)
+\frac{r(r-1)(r-2)(r-3)}{24}\left(v^4+\pi^2v^2-4v\psi''(1)+\frac{3\pi^2}{20}\right).
\end{eqnarray}
\end{widetext}

\subsection{Fitting Parameters}

\begin{table}[ht]
\caption{Fitting Parameters for nanowire samples}
\centering 
\begin{tabular}{c c c c c c}
\hline\hline 
Sample &$ \Omega_0$ ($10^{12} s^{-1}$) & $I_c(0)$ ($\mu$A) & $T_c$ (K)& $T_q$ (K) \\ [0.5ex] 
\hline 
A & 0.8 & 11.02 & 5.70 & 0.784  \\ 
B & 0.16 & 12.01 & 5.48 & 0.781 \\
C & 0.32 & 13.27 & 5.00 & 0.814 \\
D & 0.16 & 9.38 & 5.09 & 0.87 \\
E & 0.16 & 4.25 & 3.24 & 0.521 \\
F & 0.02 & 5.46 & 4.57 & 0.711 \\ [1ex] 
\hline 
\end{tabular}
\label{nanowire table} 
\end{table}
\begin{table}[ht]
\caption{Fitting Parameters for SGS junction sample 111s}
\centering 
\begin{tabular}{c c c c c}
\hline\hline 
Gate Voltage ($V_g$) &$ \Omega_0$ ($10^{9} s^{-1}$) &
$R_N$ ($\Omega$) & $T_c$ (K)& D (cm$^2$/s)\\ [0.5ex] 
\hline 
1 & 1.5 & 138 & 7.2 & 51 \\ 
3 & 1.5 & 88 & 7.2 & 54 \\
5 & 0.2 & 72.5 & 7.2 & 57\\
-1 & 0.6 & 210 & 7.2 & 48\\ [1ex] 
\hline 
\end{tabular}
\label{graphene table} 
\end{table}
\begin{table}[ht]
\caption{Parameters of the thermal and quantum escape rates for
various degrees of Ohmic damping encoded by the quality factor $Q$
of a junction. The table is compiled from previous studies of phase
slip junction (PSJ) and Josephson junction (JJ)
models~\cite{LA,MH,LO,Grabert,Fisher,HTB,Garg,Tinkham-PRB,GZ,Zaikin-Review}.
$U_0$ stands for $U(I=0,T)$.}
\centering 
\begin{tabular}{c c c c c c}
\hline\hline 
Escape & Damping & $A$ & $B$ & $a$ & $b$ \\ [0.5ex] 
\hline 
Thermal-JJ & Low & $18\Omega_0U_0/5\pi QT$ & $U_0/T$ & $1$ & $3/2$ \\ 
Thermal-JJ & Moderate & $\Omega_0/2\pi$ & $U_0/T$ & -1/4 & 3/2 \\
Thermal-JJ & High & $\Omega_0Q/2\pi$ & $U_0/T$ & 0 & 3/2\\
Thermal-PCJ & Moderate & $\sqrt{U_0/T}$ & $U_0/T$ & 3/8 & 5/4\\
Quantum & None & $\sqrt{216U_0\Omega_0/\pi}$ & $36U_0/5\Omega_0$ & 5/8 & 5/4 \\
Quantum & High & $\sqrt{3U_0\Omega_0/Q^7}$ & $3\pi U_0/Q\Omega_0$ &
0 & 1
\\[1ex] 
\hline 
\end{tabular}
\label{graphene table} 
\end{table}
\begin{figure*}
 \includegraphics[width=10cm]{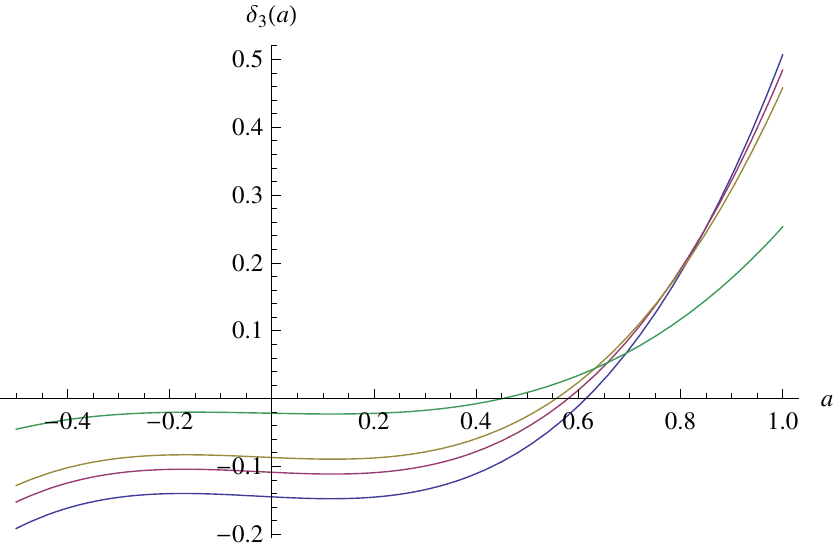}
 \caption{Correction to skewness as a function of exponent $a$ for different values of the
 ramp speed encoded by expansion parameter $Z$. $Z=15,20,25,100$ from the bottom to the top curve. Notice that
 for most models summarized in Table III parameter $a$ falls in the range where $\delta_3$ is
 nearly independent of $a$ and is determined solely by the ramp speed.
 In particular, for PSJ model $a=3/8$ and for JJ model $a=-1/4$.}
\end{figure*}

\begin{figure*}[t!]
 \includegraphics[width=15cm]{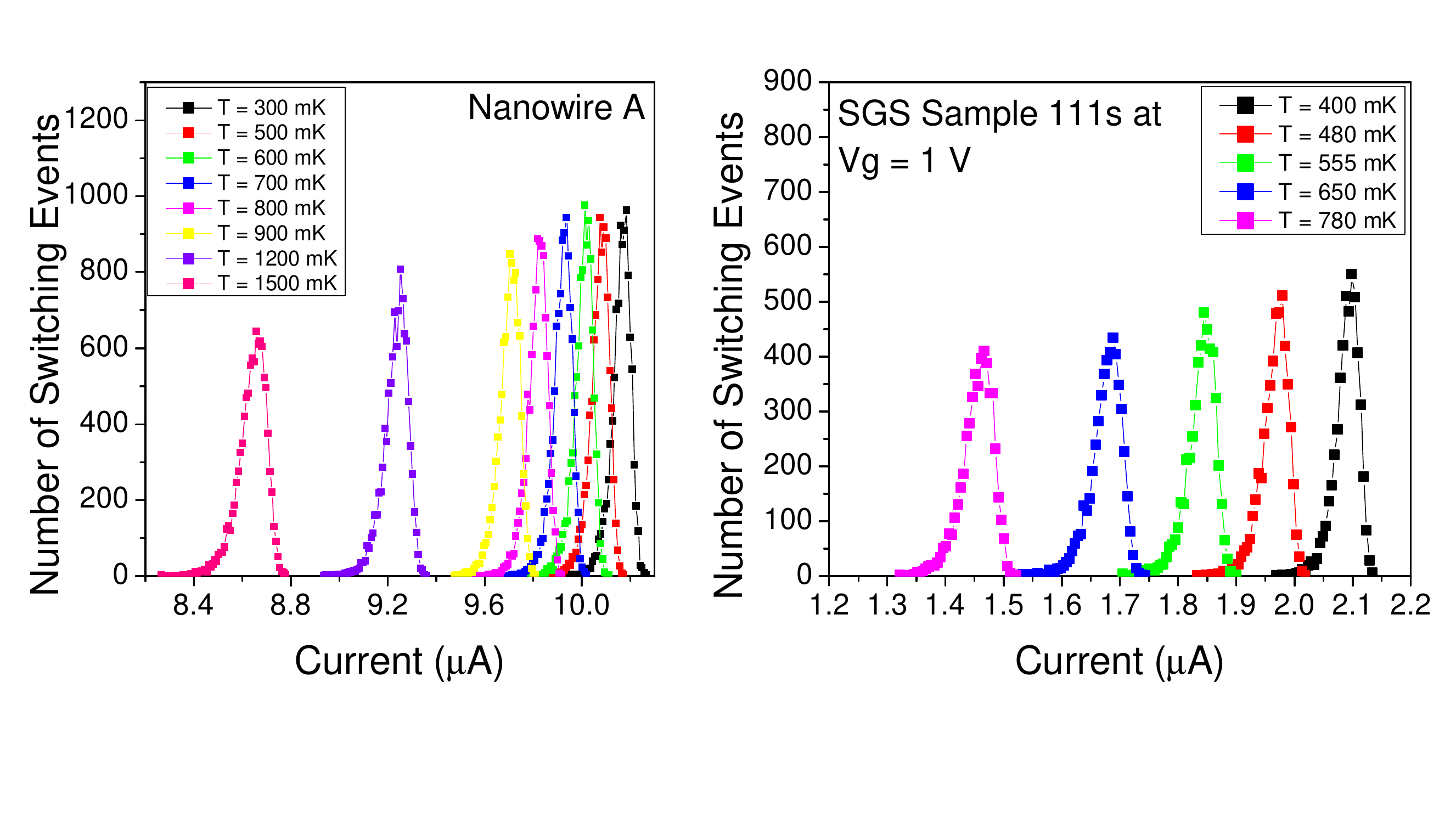}
 \caption{Number of switching events versus current for superconducting nanowire A (left)
 and for superconductor-graphene-superconductor junction, Sample 111s, at a gate voltage of $V_g=1$ V (right) plotted at various temperatures. To plot these distributions of the switching current the bin size was chosen 8.15 nA for the nanowire and 5 nA for graphene. Solid lines are given as guides to the eye. Note that the height of the distribution for the nanowire saturates below ~0.7K, indicating that the switching is initiated by quantum phase slips at these low temperatures.}\label{distributions}
\end{figure*}

\subsection{Analysis of correction terms $\delta_3$ and $\delta_4$}

For our results to be valid the ramp rate $|\dot\epsilon|$ must
be small (otherwise the system will not be able to
maintain an equilibrium from the initial state). The typical range of
the sweep rates used in the experiments is $|\dot\epsilon|\sim
(10^{-7}-10^{-5})$s$^{-1}$. At operating temperature $T\sim0.3$K the
typical value of the escape frequency translates into the parameter
$A\sim10^{-12}$s$^{-1}$. For the critical current at that
temperature we have $I_c\sim10^{-5}$A which translates into the
parameter $B\sim10^3$ that gives the scale for the height of the
potential barrier for the escape. This together leads to an estimate of
the expansion parameter $Z\sim 15$.

Now we present estimates that give some measure of the importance of
correction terms $\delta_3$ and $\delta_4$ in our expressions for
skewness and kurtosis versus power $a$ and expansion parameter $Z$
for the JJ model with $b=3/2$. Given that theoretical calculations
of the attempt frequency $\Omega$ (the pre-exponential factor in the
escape rate) are quite challenging, the power-law exponent $\nu$
(or, in the reduced notation of Eq.~6 parameter $a$) is uncertain
and is still a subject of debate. To the leading order the
expression for $\delta_3$ reads
\begin{widetext}
\begin{equation}
\delta_3=\frac{1}{60Zb^4}[90a\pi^2v(v-1)-11\pi^4(b-1)-180\psi''(1)(a-v(b-1))].
\end{equation}
\end{widetext}
From Fig.~1 we conclude that the magnitude of $\delta_3$ is about
10\% of the universal value of $S$. The expression for $\delta_4$ is
implicit in Eq.~(4) for $f_4(n,\ln Z)$ and also $f_5(n,\ln Z)$.
Taking $a=0$ for simplicity we find
\begin{equation}
\delta_4=\frac{1-b}{5Zb^5}[3\gamma\pi^4-15\pi^2\psi''(1)-10\psi^{(4)}(1)].
\end{equation}
For the value of $Z$ estimated above ($Z\sim15$) and for the JJ model with
$b=3/2$ one finds $\delta_4\approx-0.679$ which changes $K$ by about
10\%. These estimates provide a firm assurance for the consistency
of our results.

\subsection{Switching current distribution}

For completeness we provide several representative examples of
phase-slip induced switching current distributions for both types of
devices used in our experiments, see Fig.~\ref{distributions} for
details.

\subsection{Distribution moments for SGS sample 105.}

\begin{figure*}
 \includegraphics[width=11cm]{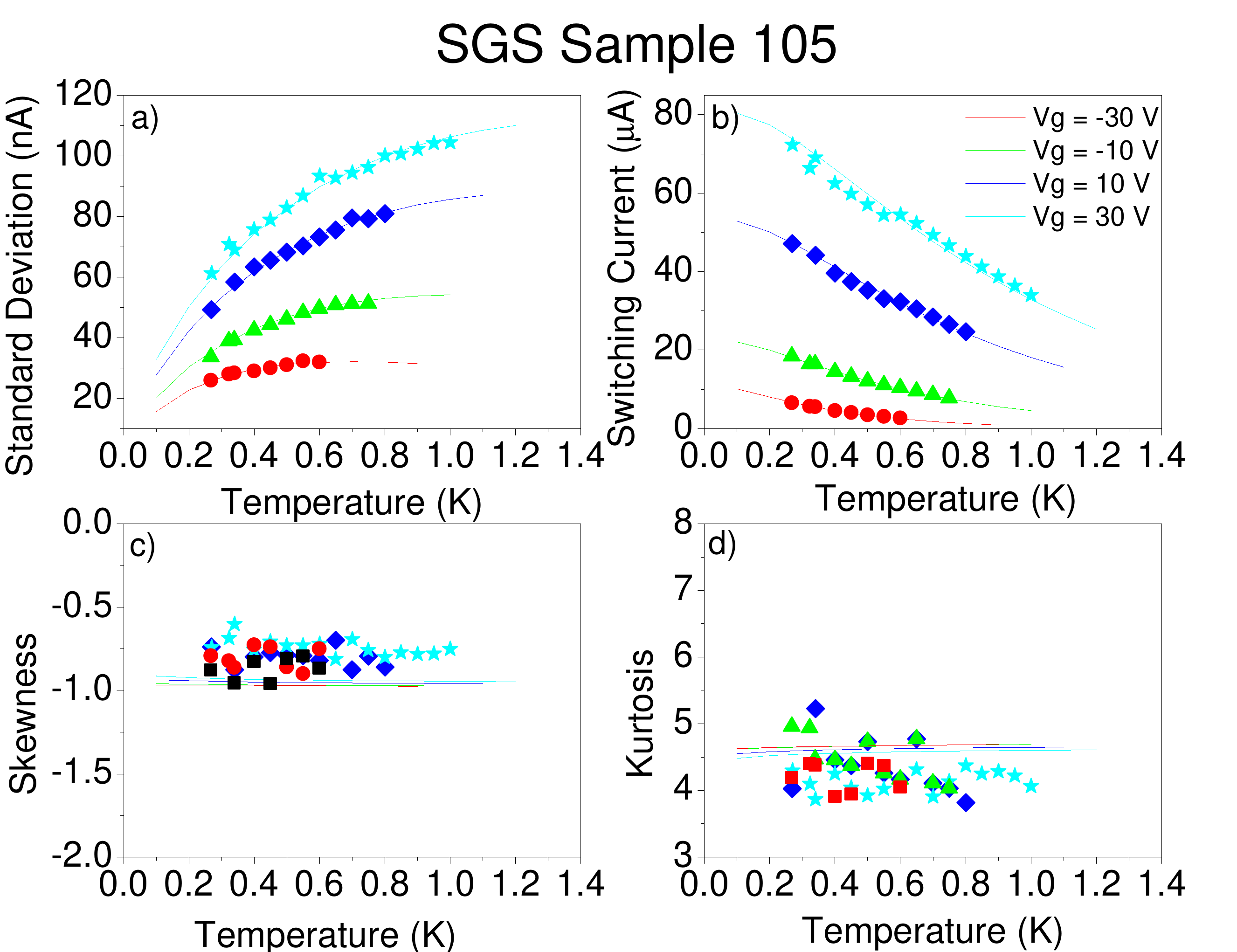}
 \caption{Mean, standard deviation, skewness and kurtosis of switching
 current distribution for SGS sample 105. Symbols correspond to gate
 voltages as indicated on the legend in the top right corner.
 Solid lines are obtained through a fitting procedure analogous
 to that utilized for sample 111s with no extraneous noise.
 Fitting parameter $R_N$ lies in the range $R_N \approx 2- 10\,
 \Omega$ while $\rm D \approx 10-30 \rm cm^2/s$.
 These values are consistent with the geometry of this sample.}
\label{SGS2}
\end{figure*}

Fig.~\ref{SGS2} shows temperature dependence of first four moments
of switching current distributions for SGS sample 105 - our second
graphene sample. Each point is based on $10^4$ measurements of the
switching current. This junction has a much larger width (dimension
across the current) $W=214\, \mu$m and hence graphene area,  but
shows qualitatively the same behavior of higher moments as SGS
sample 111s ($W=9.9\,\mu$m). Namely, the skewness is near $-1$ and
the kurtosis is near 5, independently of temperature or gate
voltage.
Since the sample area is much larger, it is somewhat more sensitive
to external perturbations. Namely, at temperatures below 350 mK some
anomalous switching events occur at unexpectedly low bias current.
Typically, out of $10^4$ measured events, two or three switching
events deviate significantly (by more than 10 standard deviations)
from the general population of the switching distribution. Such
anomalous events are very rare, and are believed to be  unrelated to
thermal (and/or quantum) fluctuations which we investigate here.
Therefore we eliminate such points from our statistical analysis.
They represent non-Gaussian extraneous perturbations of unknown
origin. As stated above, no more than 2 or 3 anomalous points, out
of $10^4$ total, are excluded. Note that at temperatures greater
than 350 mK no switching events have been excluded from the analysis
because no such anomalously low-current switching events were
observed at $T>350$~mK. It is clear from Fig.~\ref{SGS2} that the
agreement with the model is reasonably good at all temperatures
tested.

\subsection{Numerical Simulations of Extraneous Noise}

\begin{figure}
 \includegraphics[width=9cm]{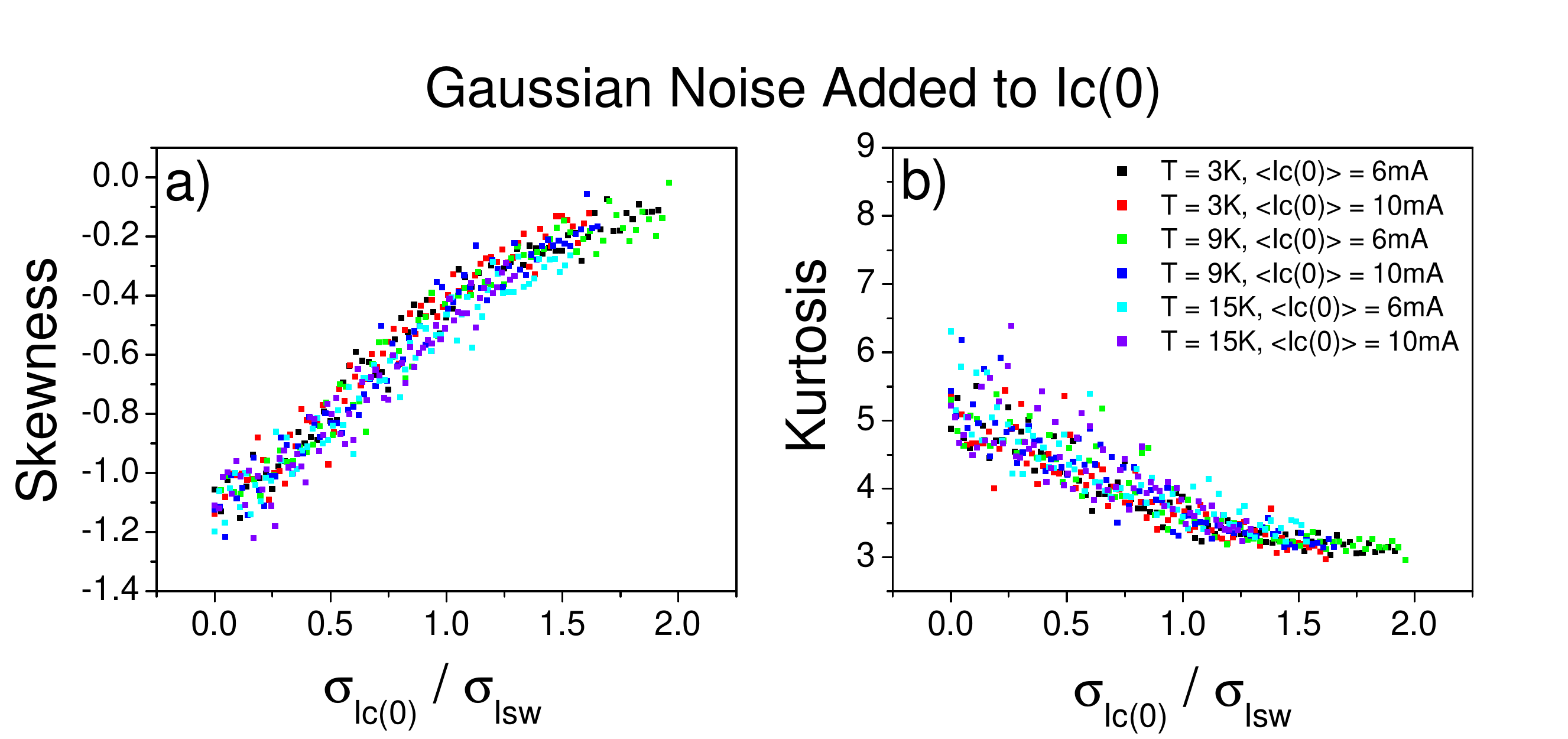}
 \caption{Numerical modeling of the variation of the skewness (a)
 and kurtosis (b) in responce to extraneous Gaussian noise
 of the critical current.
 The horizontal axis is the ratio of the standard
 deviation of the critical current of the wire $\sigma_{I_c(0)}$ and the standard
 deviation of the switching current
 $\sigma_{I_{SW}}$ which occurs if $I_c(0)$ does not fluctuate.}\label{Noise}
 \vspace{-10pt}
\end{figure}

Although most samples produced switching distributions with
$S\approx -1$ and $K\approx 5$, nanowire samples B and F and SGS sample 105 have
slight, but noticeable deviations from these values. Such deviations
pull the moments closer to those of a Gaussian distribution. We
attribute such trend to white noise present within the system which
could be caused by e.g. depairing current fluctuations brought about
by time-dependent morphological changes of our amorphous wires.
Utilizing the same numerical model as used to fit the curves in
Figs.~1 and 2 in the main text, we examined how random noise in the
critical current could affect the skewness and kurtosis of our
nanowire samples. To do this, we allowed the critical current to
vary about a mean value with Gaussian probability between the
measurement runs. Results are shown in Fig.~\ref{Noise}. We find
that upon inclusion of such extraneous noise both skewness and
kurtosis decrease in absolute values. Based on this observation we
conjecture that reduction of the magnitude of higher moments from
their universal values should be taken with caution and may signal
about the presence of undesirable noise effects.


\vspace{-10pt}

\begin{thebibliography}{00}

\bibitem{Review}
A.~A.~Clerk \textit{et al.}, Rev. Mod. Phys. \textbf{82}, 1155
(2010).

\bibitem{Nazarov}
\textit{Quantum Noise in Mesoscopic Physics}, Proceedings of the
NATO Advanced Research Workshop edited by Yu.~V.~Nazarov (Springer,
2003).

\bibitem{Prober}
B.~Reulet, J.~Senzier, and D.~E.~Prober, Phys. Rev. Lett.
\textbf{91}, 196601 (2003).

\bibitem{Devoret}
For a review, see M.~H.~Devoret \textit{et al.}, in \textit{Quantum
Tunnelling in Condensed Media}, edited by Yu.~Kagan and
A.~J.~Leggett (Elsevier, 1992), p.~313, and references therein.

\bibitem{Giordano}
N.~Giordano, Phys. Rev. Lett. 61, \textbf{2137} (1988); Phys. Rev.
Lett. 63, \textbf{2417} (1989); Phys. Rev. B 41, \textbf{6350}
(1990).

\bibitem{Bezryadin-PRL01}
C.~N.~Lau \textit{et al.}, Phys. Rev. Lett. \textbf{87}, 217003
(2001).

\bibitem{Sahu-NP09}
M.~Sahu \textit{et al.}, Nat. Phys. \textbf{5}, 503 (2009).

\bibitem{Little}
W.~A.~Little, Phys. Rev. \textbf{156}, 396 (1967).

\bibitem{Aref-RPB12}
T.~Aref \textit{et al.}, Phys. Rev. B \textbf{86}, 024507 (2012).

\bibitem{Bezryadin-Book}
A.~Bezryadin, \textit{Superconductivity in Nanowires}, (Wiley-VCH, 2012).

\bibitem{CL-model}
A.~O.~Caldeira and A.~J.~Leggett, Ann. Phys. (N.Y.) \textbf{149},
374 (1983).

\bibitem{Grabert}
H.~Grabert and U.~Weiss, Phys. Rev. Lett \textbf{54}, 1605 (1985).

\bibitem{Fisher}
M.~P.~A.~Fisher and A.~T.~Dorsey, Phys. Rev. Lett \textbf{54}, 1609
(1985).

\bibitem{GZ}
D.~S.~Golubev and A.~D.~Zaikin, Phys. Rev. B \textbf{64}, 014504
(2001); Phys. Rev. B \textbf{78}, 144502 (2008).

\bibitem{Zaikin-Review}
K.~Yu.~Arutyunov, D.~S.~Golubev, and A.~D.~Zaikin, Phys. Rep.
\textbf{464}, 1 (2008).

\bibitem{AL-PRB07}
A.~Levchenko and A.~Kamenev, Phys. Rev. B \textbf{76}, 094518
(2007).

\bibitem{Gleb-PRL11}
P.~Li \textit{et al.}, Phys. Rev. Lett. \textbf{107}, 137004 (2011).

\bibitem{Shah-PRL07}
N.~Shah, D.~Pekker, and P.~M.~Goldbart, Phys. Rev. Lett.
\textbf{101}, 207001 (2007).

\bibitem{Pekker-PRB09}
D.~Pekker \textit{et al.}, Phys. Rev. B \textbf{80}, 214525 (2009).

\bibitem{Kurkijarvi}
J.~Kurkij$\ddot{a}$rvi, Phys. Rev. B {\bf 6}, 832 (1972).

\bibitem{Bezryadin-Nature-2000}
A.~Bezryadin, C.~N.~Lau, and M.~Tinkham, Nature \textbf{404}, 971
(2000).

\bibitem{Novoselov}
K.~S.~Novoselov, \textit{et. al.}, Nature \textbf{438}, 197 (2005).

\bibitem{SM}
See Supplemenatry Material for details.

\bibitem{Lee-PRL11}
Gil-Ho Lee \textit{et al.}, Phys. Rev. Lett. \textbf{107}, 146605
(2011).

\bibitem{Tinkham}
M.~Tinkham, \textit{Introduction to Superconductivity}, 2nd ed.
(McGraw, NY, 1996).

\bibitem{Note-1}
Note that for graphene device we have set $T_q=0$ since no
low-temperature saturation in the standard deviation of the
switching current was observed on them.

\bibitem{Coskun-PRL2012}
U.~C.~Coskun \emph{et al}., Phys. Rev. Lett \textbf{108}, 097003
(2012).

\bibitem{Note-2}
The power exponent $\eta=3/2$ is characteristic for the Kramers-type
escape problem from the potential barrier discribed by the cubic
parabola.

\bibitem{ZZ}
A.~D.~Zaikin and G.~F.~Zharkov, Sov. J. Low Temp. Phys. \textbf{7}, 184 (1981).

\bibitem{Dubos-PRB2001}
P.~Dubos \emph{et al}., Phys. Rev. B \textbf{63}, 064502 (2001).

\bibitem{AL-PRB06}
A.~Levchenko, A.~Kamenev, and L.~Glazman Phys. Rev. B \textbf{74},
212509 (2006).

\bibitem{Titov}
M.~Titov and C.~W.~J.~Beenakker Phys. Rev. B \textbf{74}, 041401
(2006).

\bibitem{SGS-1}
H.~B.~Heersche \textit{et al.}, Nature (London) \textbf{446}, 56
(2007).

\bibitem{SGS-2}
F.~Miao \textit{et al.}, Science \textbf{317}, 1530 (2007).

\bibitem{SGS-3}
X.~Du, I.~Skachko, and E.~Y.~Andrei, Phys. Rev. B \textbf{77},
184507 (2008).

\bibitem{SGS-4}
C.~M.~Ojeda-Aristizabal \textit{et al.}, Phys. Rev. B \textbf{79},
165436 (2009).

\bibitem{SGS-5}
D.~Jeong \textit{et al.}, Phys. Rev. B \textbf{83}, 094503 (2011).

\bibitem{SGS-6}
I.~V.~Borzenets \textit{et al.}, Phys. Rev. Lett. \textbf{107},
137005 (2011).

\bibitem{LA}
J.~S.~Langer and V.~Ambegaokar, Phys. Rev. \textbf{164}, 498 (1967).

\bibitem{MH}
D.~E.~McCumber and B.~I.~Halperin, Phys. Rev. B \textbf{1}, 1054
(1970).

\bibitem{Tinkham-PRB}
M.~Tinkham \textit{et al.}, Phys. Rev. B \textbf{68}, 134515 (2003).

\bibitem{Bardeentest}
M.~W.~Brenner \textit{et al.},  Phys. Rev. B \textbf{83}, 184503
(2011); Phys. Rev. B \textbf{85}, 224507 (2012).

\bibitem{Bardeen}
J.~Bardeen, Rev. Mod. Phys. \textbf{34}, 667 (1962).

\bibitem{Note-3}
Fitting curves for nanowire samples were made using the cubic potential
model for $U$, but the value of $\kappa$ was adjusted, up to ~30\%
of the theoretical value, to achieve best-fits.

\bibitem{Garg-PRB1995}
A.~Garg, Phys. Rev. B \textbf{51}, 15592 (1995).

\bibitem{Note-4}
Notice that in expression (\ref{m2}) for $\sigma^2$ coefficient of
$1/Z$ term in the brackets is different from that given by Eq.~(8)
of Ref.~\onlinecite{Garg-PRB1995} which appears to be in error.

\end{thebibliography}

\begin{thebibliography}{00}

\bibitem{LA}
J.~S.~Langer and V.~Ambegaokar, Phys. Rev. \textbf{164}, 498 (1967).

\bibitem{MH}
D.~E.~McCumber and B.~I.~Halperin, Phys. Rev. B \textbf{1}, 1054
(1970).

\bibitem{LO}
A.~I.~Larkin and Yu.~N.~Ovchinnikov, Zh. Eksp. Teor. Fiz.
\textbf{86}, 719 (1984) [Sov. Phys. JETP \textbf{59}, 420 (1984)].

\bibitem{Grabert}
H.~Grabert and U.~Weiss, Phys. Rev. Lett \textbf{54}, 1605 (1985).

\bibitem{Fisher}
M.~P.~A.~Fisher and A.~T.~Dorsey, Phys. Rev. Lett \textbf{54}, 1609
(1985).

\bibitem{HTB}
P.~Hanngi, P.~Talkner and M.~Borkovec, Rev. Mod. Phys. \textbf{62},
251 (1990).

\bibitem{Garg}
A.~Garg, Phys. Rev. B \textbf{51}, 15592 (1995).

\bibitem{Tinkham-PRB}
M.~Tinkham \textit{et al.}, Phys. Rev. B \textbf{68}, 134515 (2003).

\bibitem{GZ}
D.~S.~Golubev and A.~D.~Zaikin, Phys. Rev. B \textbf{64}, 014504
(2001); Phys. Rev. B \textbf{78}, 144502 (2008).

\bibitem{Zaikin-Review}
K.~Yu.~Arutyunov, D.~S.~Golubev, and A.~D.~Zaikin, Phys. Rep.
\textbf{464}, 1 (2008).

\end{thebibliography}
\end{document}